\documentclass[%
 reprint,
 amsmath,amssymb,
 aps,
prb,
]{revtex4-1}
\usepackage[dvipdfmx]{graphicx}% Include figure files
\usepackage{dcolumn}% Align table columns on decimal point
\usepackage{bm}% bold math
\usepackage{braket}
\begin{document}

\preprint{APS/123-QED}

\title{Spin-Orbit Interaction Induced in Graphene by Transition-Metal Dichalcogenides}% Force line breaks with \\

\author{T. Wakamura$^1$, F. Reale$^2$, P. Palczynski$^2$, M. Q. Zhao$^3$, A. T. C. Johnson$^3$, S. Gu\'{e}ron$^1$, C. Mattevi$^2$, A. Ouerghi$^4$}
\author{H. Bouchiat$^1$}%
 \email{helene.bouchiat@u-psud.fr}
\affiliation{$^1$Laboratoire de Physique des Solides, Universite Paris-Sud, 91400, Orsay, France}%

\affiliation{$^2$Department of Materials, Imperial College London, Exhibition Road, London, SW7 2AZ, United Kingdom}%

\affiliation{$^3$Department of Physics and Astronomy, University of Pennsylvania, 209S 33rd Street, Philadelphia, Pennsylvania 19104 6396, United States}

\affiliation{$^4$Centre de Nanosciences et de Nanotechnologies, CNRS, University of Paris-Sud, Universite Paris-Saclay, C2N, Marcoussis 91460, France}
%\affiliation{
 %Third institution, the second for Charlie Author
%}%
%\author{Helene Bouchiat}
%\affiliation{%


\date{\today}% It is always \today, today,

\begin{abstract}
We report a systematic study on strong enhancement of spin-orbit interaction (SOI) in graphene induced by transition-metal dichalcogenides (TMDs). Low temperature magnetotoransport measurements of graphene proximitized to different TMDs (monolayer and bulk WSe$_2$, WS$_2$ and monolayer MoS$_2$) all exhibit weak antilocalization peaks, a signature of strong SOI induced in graphene. The amplitudes of the induced SOI are different for different materials and thickness, and we find that monolayer WSe$_2$ and WS$_2$ can induce much stronger SOI than bulk WSe$_2$, WS$_2$ and monolayer MoS$_2$. The estimated spin-orbit (SO) scattering strength for graphene/monolayer WSe$_2$ and graphene/monolayer WS$_2$ reaches $\sim$ 10 meV whereas for graphene/bulk WSe$_2$, graphene/bulk WS$_2$ and graphene/monolayer MoS$_2$ it is around 1 meV or less. We also discuss the symmetry and type of the induced SOI in detail, especially focusing on the identification of intrinsic (Kane-Mele) and valley-Zeeman (VZ) SOI by determining the dominant spin relaxation mechanism. Our findings pave the way for realizing the quantum spin Hall (QSH) state in graphene. 
%\begin{description}
%\end{description}
\end{abstract}

\pacs{Valid PACS appear here}% PACS, the Physics and Astronomy
\maketitle
\section{\label{sec:level1}Introduction\protect\\ }
%\tableofcontents
Two dimensional (2D) layered materials have provoked tremendous interest since the first demonstration of mechanical exfoliation of graphene \cite{geim1}. A growing number of recent reports on these materials have revealed that they exhibit many intriguing physical properties, including superconductivity\cite{super1, super2, super3, super4}, ferromagnetism \cite{ferro1, ferro2, ferro3}, quantum spin Hall (QSH) state \cite{qsh1, qsh2, qsh3, qsh4} and Weyl semimetal state \cite{weyl1, weyl2, weyl3}. While these properties in the two dimensional limit are of great interest in themselves, one can also exploit them as building blocks to generate novel phenomena through interface interactions between two different materials \cite{geim2}.  

Among the many intriguing phenomena realized via interface interactions, inducing spin-orbit interaction (SOI) in graphene is particularly attractive for application to spintronics \cite{manuel} and topological physics \cite{kane1, kane2}. So far many works have reported both theoretically and experimentally that strong SOI can be induced in graphene extrinsically by hydrogenation \cite{balakrishnan, gmitra1, castroneto}, adatom deposition \cite{adatom1, wu, balakrishnan2} and intercalation of heavy atoms between graphene and metallic substrates in chemical vapor deposition (CVD)-grown graphene \cite{adatom2, adatom3, adatom4, gold}. On the other hand recent works have focused on graphene/transition-metal dichalcogenides (TMDs) heterostructures as a platform. TMDs are 2D materials like graphene, but their SOI is much larger than that of graphene owing to the heavy transition metals \cite{xiao}. This method is more advantageous than the previous ones since it preserves the quality of graphene. Recent theoretical and experimental studies revealed that graphene proximitized by TMD can acquire strong SOI through interfacial coupling \cite{avsar2, wang, wang2, yang, yang2, gmitra2, gmitra3, roche, cysne, offdani, ilic, Alsharari, valenzuera, vanwees, frank, zihlmann, wakamura, eroms1, eroms2}. Experimentally, the induced SOI is estimated through transport measurements of nonlocal voltages induced by spin currents or weak antilocalization (WAL) measurements at low temperatures, exhibiting strongly enhanced spin-orbit scattering in graphene. However, estimates of the induced SOI are not in good agreement with the theoretically calculated values based on \textit{ab initio} calculations, and there are sometimes one order of magnitude differences between them.

Beyond the amplitudes, the symmetry of the induced SOI is important. Indeed, the QSH state is one of the intriguing states expected to emerge if strong enough SOI is induced in graphene. It requires $z \rightarrow -z$ symmetric SOI, where the $z$ axis is normal to the graphene plane \cite{kane1, kane2}. For pristine graphene this symmetric SOI is provided by intrinsic (Kane-Mele (KM)) SOI. On the other hand $z \rightarrow -z$ asymmetric SOI is also expected in realistic experimental situations, induced by Rashba SOI due to the inversion symmetry breaking by the substrate or perpendicular electric fields.
\begin{figure*}[t]
\begin{center}
\includegraphics[width=17cm,clip]{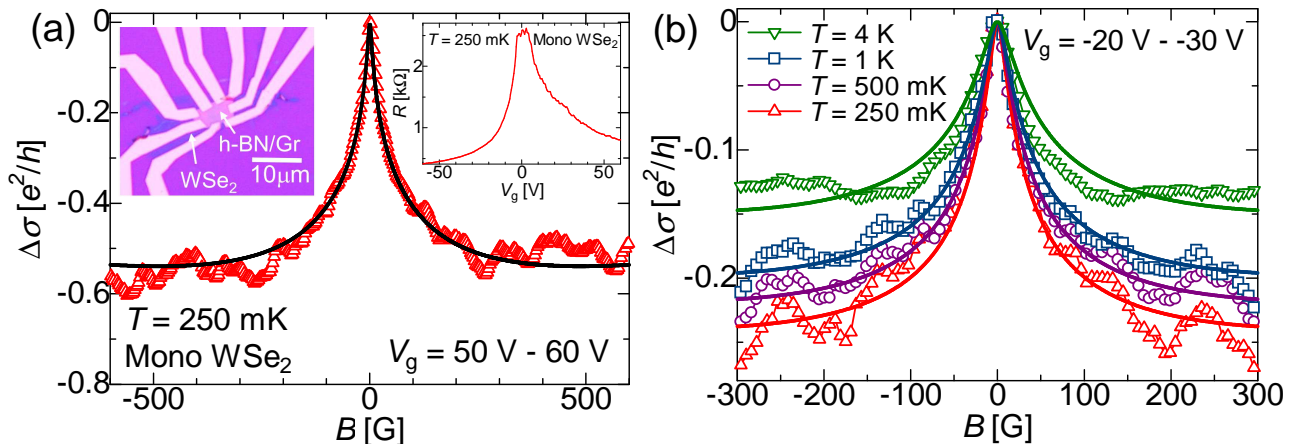}
\caption{(a) Magnetoconductivity correction ($\Delta \sigma (B) \equiv \sigma (B) - \sigma(0)$) for Mono WSe$_2$ averaged over 50 curves corresponding to 50 values of $V_g$ between 50 V and 60 V at 250 mK. A clear WAL peak and flat tails for higher $B$ regions are observed. The solid curve represents the fit based on the theoretical formula (1). The left inset shows the optical microscope image of the Mono WSe$_2$ device. The gate voltage dependence of resistance is displayed in the right inset. (b) $\Delta \sigma (B)$ curves at different temperatures, averaged over 50 curves with $V_g$ between $-20$ V and $-30$ V. Similar tendency as in (a) can be seen in the shape of each curve. The solid lines are theoretical fits.
.} 
\label{fig1}
\end{center}
\end{figure*} 
Recent theoretical studies propose the existence of a new type of SOI in graphene/TMD systems: a valley-Zeeman (VZ) SOI induced in graphene due to the broken sublattice symmetry \cite{roche, zihlmann}. This SOI provokes the spin splitting of degenerate bands, with out-of-plane spin polarization at $K$ and $K'$ points, and an opposite spin-splitting in different valleys. Analogous to the Zeeman-splitting, the SOI is named VZ SOI because the effective Zeeman fields are valley-dependent. This is the dominant SOI in TMDs, and it is also predicted to be induced in graphene on TMD \cite{roche}. One of the important consequences of the VZ SOI is the anisotropic spin relaxation, as revealed by recent first principle calculation and experimental studies \cite{roche, zihlmann, valenzuera, vanwees}. In terms of symmetry the VZ SOI is $z \rightarrow -z$ symmetric, and it is predicted to yield topologically-unprotected edge states \cite{frank}.          

In this paper, we present a systematic study of SOI induced in graphene/TMD heterostructures. 
We measure magnetoresistance at low temperatures and demonstrate that strong SOI is induced in graphene by all investigated TMD crystals: monolayers of WSe$_2$, WS$_2$ and MoS$_2$, as well as bulk WSe$_2$ and WS$_2$. We observe a clear difference between monolayer and bulk TMDs in the capacity to induce strong SOI in graphene as reported before \cite{wakamura}. For both WSe$_2$ and WS$_2$, the induced SOI is stronger when the TMD is monolayer than bulk. Monolayer WSe$_2$ and WS$_2$ induce comparable amplitudes of  SOI, whereas monolayer MoS$_2$ generates much smaller SOI. For all samples with different TMDs, we find that the $z \rightarrow -z$ symmetric SOI is dominant. To identify the type of $z \rightarrow -z$ symmetric SOI, we elucidate the spin relaxation mechanism by plotting the relation between the momentum relaxation time $\tau_p$ and the spin-orbit time $\tau_{\rm so}$. The dominant Elliot-Yafet (EY) mechanism around the Dirac point indicates the importance of KM SOI in this  low doping region. We also discuss the possibility of VZ SOI and the reason for the suppressed Rashba SOI, taking into account recent reports on band structures of graphene/TMD systems measured by angle-resolved photoemission spectroscopy (ARPES) \cite{arpes1, arpes2, arpes3} and scanning tunneling microscopy (STM) studies on Moir\'e patterns \cite{moire1, moire2, moire3}.
\section{\label{sec:level1}Weak Antilocalization in graphene\protect\\ }
\begin{figure*}
\begin{center}
\includegraphics[width=17cm,clip]{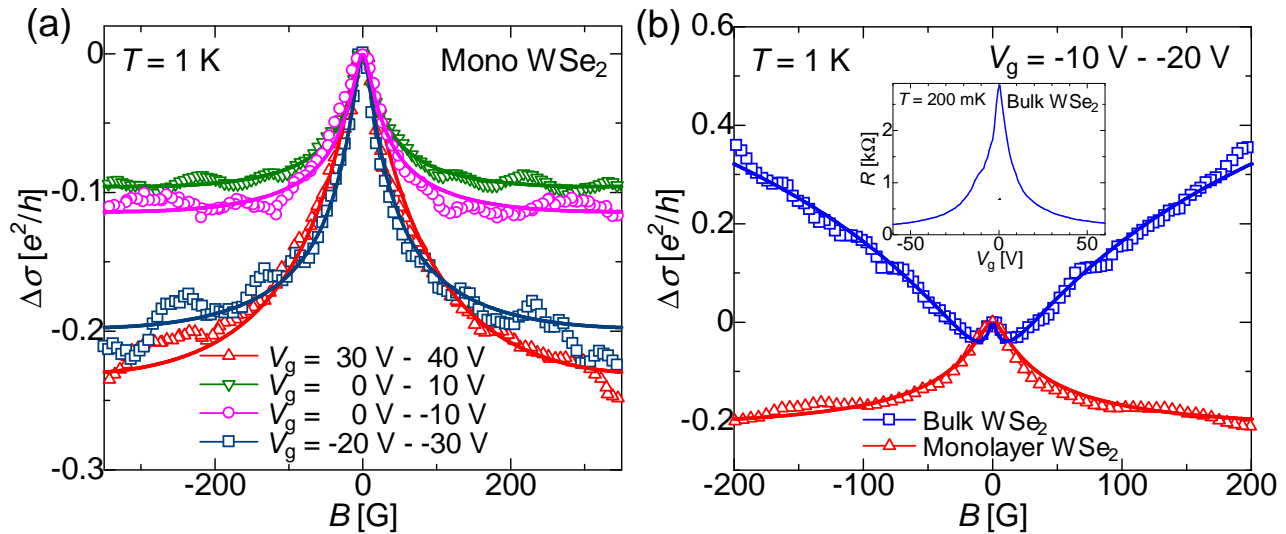}
\caption{$\Delta \sigma (B)$ curves for Mono WSe$_2$ for different gate voltage ranges at 1 K.  Clear WAL peaks are observed in all $V_g$ ranges. (b) Comparison of $\Delta \sigma (B)$ curves from Mono WSe$_2$ and Bulk WSe$_2$ in the same $V_g$ range. While the curve for Mono WSe$_2$ is characterized by flat tails for high $B$ region, that of Bulk WSe$_2$ exhibit instead a striking upturn with magnetic field following the small WAL peak around $B$ = 0. This demonstrates that stronger SOI  is induced in graphene for Mono WSe$_2$ than for Bulk WSe$_2$. In the inset we show $R$ vs $V_g$ curve for Bulk WSe$_2$.}
\label{Figure2}
\end{center}
\end{figure*}
We exploit weak (anti)localization measurements to estimate the SOI induced in graphene by TMDs. At low temperatures, large coherence length of electrons causes quantum interference between time-reversed pairs of closed trajectories of electron wave packets, leading to weak localization (WL) of electrons \cite{bergmann}. When SOI is sufficiently strong, the spin of the electrons rotates during a closed loop, leading to an additional $\pi$ phase difference between the time-reversed pairs of the electron wave packets. This results in antilocalization of electrons (WAL effect). An external magnetic field breaks  time-reversal symmetry, and as a result when WL (WAL) is dominant the resistance decreases (increases) with an increasing magnetic field. Thus magnetoresistance measurements allow to identify the regime (WL or WAL) to which the system belongs, and  provide an estimate of the SOI amplitude.

Dirac fermions in graphene are known to be robust against disorder due to the chiral nature of valley-conserving transport \cite{ando}. However, short range elastic scattering gives rise to intervalley scattering. When the intervalley scattering rate $\tau_{iv}^{-1}$ is large compared to the dephasing rate ($\tau_\phi^{-1}$), \cite{WL1, WL2} localization of Dirac fermions in graphene is restored. Many experimental studies have shown evidences of  weak localization at low temperatures where $\tau_{iv} \ll \tau_\phi$ \cite{WL1, WL2, WL3, WL4}. In this regime, if graphene acquires strong SOI it is therefore possible to observe weak antilocalization (WAL) effects due to the real spin-orbit coupling rather than the pseudospin-orbit coupling \cite{Tikhonenko}. The theoretical formula for the magnetoconductivity correction in the WAL regime for $\tau_{iv} \ll \tau_\phi$ is written as \cite{mccann}
\begin{multline}
\Delta \sigma (B) = -\frac{e^2}{2 \pi h} \left[F \left( \frac{\tau_B^{-1}}{\tau_\phi^{-1}} \right) - F \left( \frac{\tau_B^{-1}}{\tau_\phi^{-1} + 2 \tau_{\rm asy}^{-1}} \right) \right.\\
\left. -2 F \left( \frac{\tau_B^{-1}}{\tau_\phi^{-1} + \tau_{\rm so}^{-1}} \right) \right],
\label{eq1}
\end{multline}
where $F(x) = \ln(x)+\psi(1/2 + 1/x)$, with $\psi(x)$ the digamma function, $\tau_{\rm so}^{-1} = \tau_{\rm sym}^{-1} + \tau_{\rm asy}^{-1}$, where sym (asy) denotes the symmetric (asymmetric) contribution to the SOI (discussed below in detail) and $\tau_B = \hbar/4eDB$ with $D$ the diffusion coefficient. Fits of this formula to the experimental magnetoconductance curves provide three parameters $\tau_\phi$, $\tau_{\rm asy}$ and $\tau_{\rm so}$. $\tau_{\rm so}$ determines the total amplitude of SOI in the system, and the $\tau_{\rm sym}$ ($\tau_{\rm asy}$) term is associated with $z \rightarrow -z$ symmetric (asymmetric) SOI. In the case of graphene on TMD, $z \rightarrow -z$ symmetric SOI includes KM and VZ SOI, and the $z \rightarrow -z$ asymmetric SOI is attributed to Rashba or pseudospin inversion asymmetry (PIA) SOI. Details of these different types of SOI will be discussed in a later section. As predicted in the original papers by Kane and Mele \cite{kane1, kane2}, to induce the QSH state, dominant KM SOI and small Rashba SOI are required to conserve $s_z$ as a well-defined quantum number. Therefore, if other types of SOI are not considered (e.g. VZ SOI), the ratio between $\tau_{\rm sym}$ and $\tau_{\rm asy}$ is a key factor to determine the possibility to realize the QSH state in the system.
\begin{table}[b]%The best place to locate the table environment is directly after its first reference in text
\caption{\label{tab:table3} Summary of the mobility and size of each sample. }
\begin{ruledtabular}
\begin{tabular}{lcc}
\textrm{Sample}&
\textrm{Mobility [cm$^2$V$^{-1}$s$^{-1}$]}&
\textrm{Size [$\mu$m $\times$ $\mu$m]}\\
\colrule
Mono WSe$_2$ & 21000 & 6$\times$8\\
Mono WS$_2$ A & 12000 & 6$\times$12\\
Mono WS$_2$ B & 7000 & 5$\times$6\\
Mono MoS$_2$ A & 4700 & 6$\times$12\\
Mono MoS$_2$ B & 1850 & 7$\times$8\\
Bulk WSe$_2$ & 21600 & 5$\times$7\\
Bulk WS$_2$ A & 9000 & 5$\times$8\\
Bulk WS$_2$ B & 7000 & 5$\times$5\\
\end{tabular}
\end{ruledtabular}
\end{table}
\section{\label{sec:level1}Sample Preparation and Measurement Details\protect\\ }
\begin{figure*}
\begin{center}
\includegraphics[width=17cm,clip]{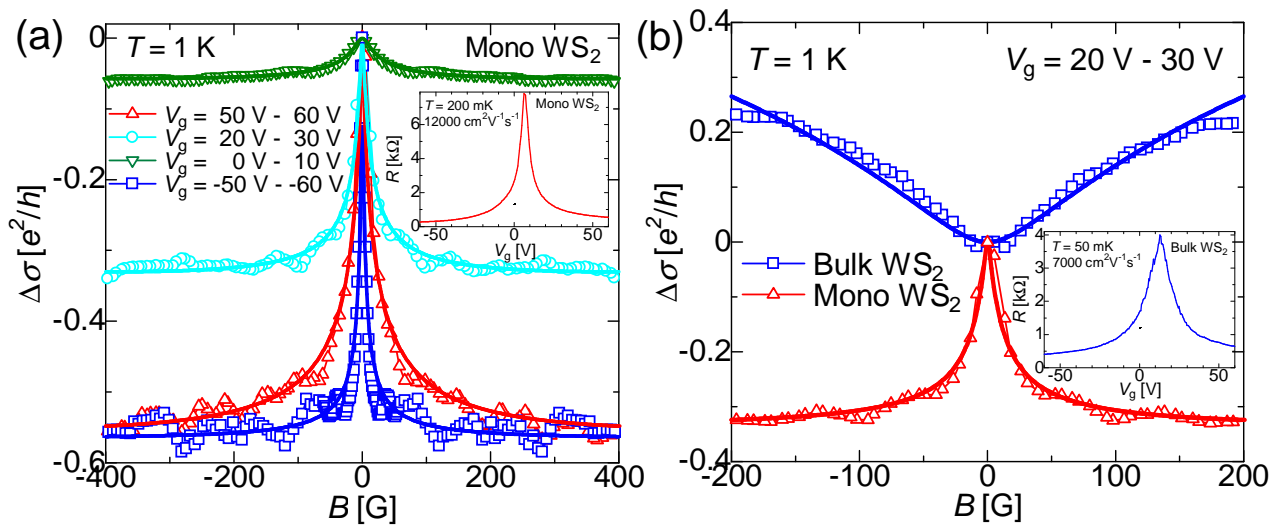}
\caption{(a) $\Delta \sigma (B)$ for different gate voltage ranges obtained from Mono WS$_2$ A at 1 K. The clear WAL peaks demonstrate the strong induced SOI. Interestingly, the peak is sharper for $-$50 V $<$ $V_g$ $<$ $-$60 V than for 50 V $<$ $V_g$ $<$ 60 V. All curves exhibit flat tails for high $B$ range. The solid lines are theoretical fits. The inset shows the $V_g$ dependence of $R$. (b) Comparison of $\Delta \sigma (B)$ between Mono WS$_2$ A and Bulk WS$_2$ B. As seen for the graphene/WSe$_2$ samples, Bulk WS$_2$ shows upturns following the WAL peak. The inset displays $V_g$ dependence of $R$. Theoretical fits are shown by the solid lines.}
\label{fig1}
\end{center}
\end{figure*}
Graphene for all heterostructures is prepared by mechanical exfoliation of natural graphite on SiO$_2$(285 nm thickness)/doped-Si substrates. Monolayer graphene is identified by means of optical contrast under the microscope, and by quantum Hall effect measurements. For TMDs, the fabrication process is different for different materials: Monolayer WS$_2$ and MoS$_2$ are prepared by the CVD method \cite{wakamura, arpes2, arpes3, reale}, and the other TMDs are prepared by mechanical exfoliation  of bulk cristals on SiO$_2$/doped-Si substrates. For graphene/monolayer TMD samples, graphene is transferred on a monolayer TMD by using polymethyl methacrylate (PMMA) or mechanically exfoliated hexagonal boron-nitride (hBN) using polydimethylsiloxane (PDMS). For graphene/bulk TMD samples, a bulk TMD is deposited on graphene. Conventional electron beam lithography techniques are employed to fabricate electrical contacts, and 5 nm Ti and 100 nm Au are subsequently deposited by electron gun evaporation. Measurements are performed in a dilution refrigerator employing a conventional lock-in technique with an excitation current of $I_{\rm ac}$ = 10 nA and 77 Hz unless otherwise noted. In the following sections, for simplicity graphene/monolayer TMD structures are termed Mono MX$_2$ (M= W or Mo, X= S or Se) and graphene/bulk ones Bulk MX$_2$. In Table \ref{tab:table3} we give the geometry and mobility of all investigated samples.
%Summary of the fabrication details are given in the Supplemental Material \cite{suppl}.
\section{\label{sec:level1}Experimental Results\protect\\ }

We evaluate the SOI induced in graphene on various TMDs by means of magnetotransport measurements at sufficiently low temperatures, where the WAL due to the chirality of graphene is negligible because $\tau_{iv} \ll \tau_\phi$ \cite{Tikhonenko}. This point will be discussed further in the section VC. To obtain clear weak (anti)localization peaks, we average over 50 curves with different $V_g$ in a 10 V window. This is because the height of the peaks is of the order of $e^2/h$, the same order of magnitude as universal conductance fluctuations (UCFs) since the sample size is comparable to the phase coherence length. In the following subsections we show the experimental results of the magnetotransport measurements for each graphene/TMD heterostructure. In the WAL data we obtain the magnetoconductivity correction ($\Delta \sigma(B)$) by converting the original data of two-terminal resistance ($R=V/I$) with subtraction of contact resistance and taking into account the aspect-ratio of the device.
\subsection{\label{sec:level2}Graphene/WSe$_2$ structures}
In this subsection we show the experimental results obtained from Mono WSe$_2$ and Bulk WSe$_2$. WSe$_2$ has the largest intrinsic SOI among TMDs both in the valence band and conduction band \cite{xiao, kosmider, burkard}. 

We first discuss the results for Mono WSe$_2$. The inset of Fig. 1(a) shows the gate voltage ($V_g$) dependence of the resistance at 250 mK for Mono WSe$_2$\cite{gmitra3}, and the optical microscope image of the device. The resistance exhibits slight asymmetry in $V_g$, and an anomalous saturation is observed around the Dirac point. The origin of this plateau is still unclear. We note that the resistivity of the monolayer WSe$_2$ is much larger than that of graphene, thus the charge transport is dominated by graphene, as evidenced by the $R(V_g)$ curve that is typical of graphene. The calculated mobility from the inset of Fig. 1(a) is 21000 cm$^2$V$^{-1}$s$^{-1}$. This mobility is higher than that of our previous study of graphene on WS$_2$ \cite{wakamura} but lower than other reports \cite{zihlmann, eroms1} 

Figures 1(a) and (b) show the conductivity correction ($\Delta \sigma \equiv \sigma(B) - \sigma(0)$) as a function of the magnetic field ($B$) applied perpendicular to the graphene plane, in the window of $V_g$ specified in the figure. For all temperatures between 250 mK and 4 K, $\Delta \sigma (B)$ exhibits WAL with a clear peak at $B$ = 0, indicating strong SOI induced in graphene. Similar peaks are observed for all the gate voltage ranges between $-$60 V and 60 V. In Fig. 2(a) we compare representative $\Delta \sigma(B)$ curves for different gate voltage ranges. The $\Delta \sigma (B)$ curves have a similar shape for the electron-doped and hole-doped region.

It is interesting that outside the central peaks, all $\Delta \sigma(B)$ curves are flat with increasing $B$. As pointed out in the previous studies \cite{wakamura, bergmann, roche2}, the flat tails in high $B$ regions are a signature of strong induced SOI, as will be discussed further below.

We next discuss the experimental results obtained from Bulk WSe$_2$. Previous studies have pointed out striking differences in electrical and optical properties between monolayer and bulk TMDs, and among them the different band structures are especially influential for transport properties \cite{klein, rama, terrones}. %We have already reported that these differences are also reflected in the induced SOI in graphene in proximity to monolayer or bulk WS$_2$ \cite{wakamura}.
To compare the induced SOI in graphene on monolayer and bulk WSe$_2$, we therefore measure the magnetoresistance of graphene on bulk WSe$_2$. We here show the data of the sample with the mobility of 21600 cm$^2$V$^{-1}$s$^{-1}$, similar to that of Mono WSe$_2$ sample discussed above. The Dirac point of Bulk WSe$_2$ is located at $V_g$ = 0, just as for Mono WSe$_2$. Figure 2(b) displays the comparison of the $\Delta \sigma (B)$ curves taken for Mono WSe$_2$ and Bulk WSe$_2$ for the same gate voltage range. The shapes of the two curves are clearly different, and that from Bulk WSe$_2$ displays a striking upturn in the high B region, which contrasts with the flat or slightly downward sloping magnetoconductance of Mono WSe$_2$. As discussed in the next section, this upturn is the signature of moderate SOI induced in graphene, smaller than induced by monolayer WSe$_2$ and WS$_2$. Similar shape differences are also observed for other gate voltage regions.

\subsection{\label{sec:level2}Graphene/WS$_2$ structures}
Since we have already reported strong SOI induced in the graphene/WS$_2$ structures in our previous paper \cite{wakamura}, we here briefly provide an overview of the results obtained from the graphene/monolayer WS$_2$ (Mono WS$_2$) and graphene/bulk WS$_2$ (Bulk WS$_2$) samples, using data not shown in our previous paper \cite{wakamura}. Figure 3(a) shows $\Delta \sigma (B)$ from one of the Mono WS$_2$ samples for four different gate voltage ranges. The mobility of this sample is 12000 cm$^2$V$^{-1}$s$^{-1}$. For all gate voltage ranges, including close to the Dirac point, we observed WAL, a signature of the strong SOI induced in graphene by monolayer WS$_2$. In contrast to Mono WSe$_2$, the shape of $\Delta \sigma (B)$ is electron-hole asymmetric in Mono WS$_2$. We note that for the data with $V_g$ between $-$50 V and $-$60 V (shown in Fig. 3(a)) a temperature-independent background signal is subtracted from the original data (see the discussions below). We also carried out low-temperature magnetotransport measurements for Bulk WS$_2$, and one $\Delta \sigma (B)$ curve is compared with that of Mono WS$_2$. The mobility of the sample is 7000 cm$^2$V$^{-1}$s$^{-1}$. As demonstrated for the graphene/WSe$_2$ heterostructure, there is a striking difference in the shape of the curves between Mono and Bulk WS$_2$. The smaller peak around $B$ = 0 and strong upturn for higher $B$ region for the Bulk WS$_2$ samples indicate that the induced SOI is smaller for Bulk WS$_2$ than for Mono WS$_2$.  
\begin{figure}
\includegraphics[width=8cm,clip]{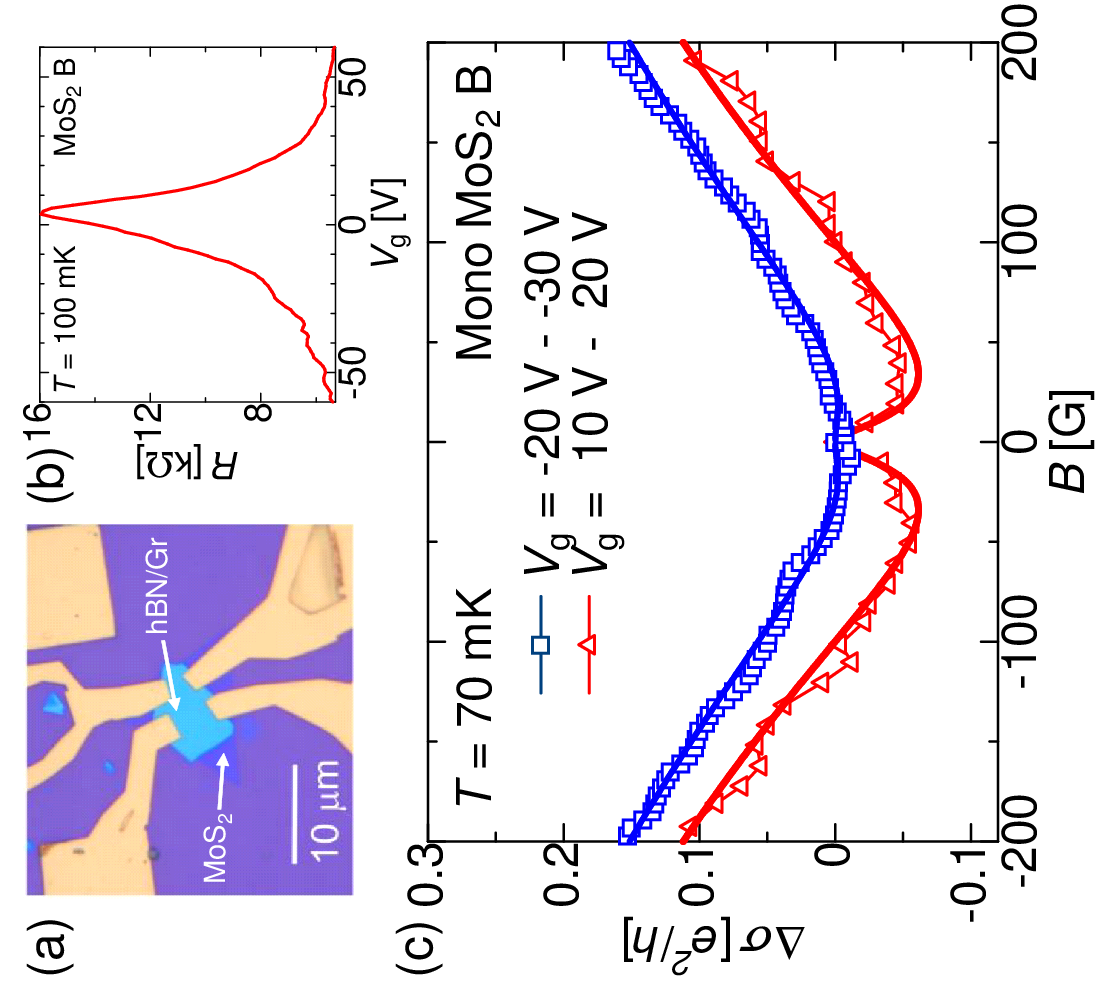}
\caption{(a) An optical microscope image of Mono MoS$_2$ B sample. Graphene is deposited on a CVD grown monolayer MoS$_2$. (b) Resistance as a function of gate voltage at 100 mK of Mono MoS$_2$ B. (c) $\Delta \sigma (B)$ for $V_g$ between  10 V and 20 V, and between -20 V and -30 V at 70 mK. In contrast to Mono WSe$_2$ and Mono WS$_2$, both curves exhibit large upturns when $B$ is large. The solid lines show theoretical fits.}
\label{fig1}
\end{figure}
\begin{figure*}[tb]
\begin{center}
\includegraphics[width=17cm,clip]{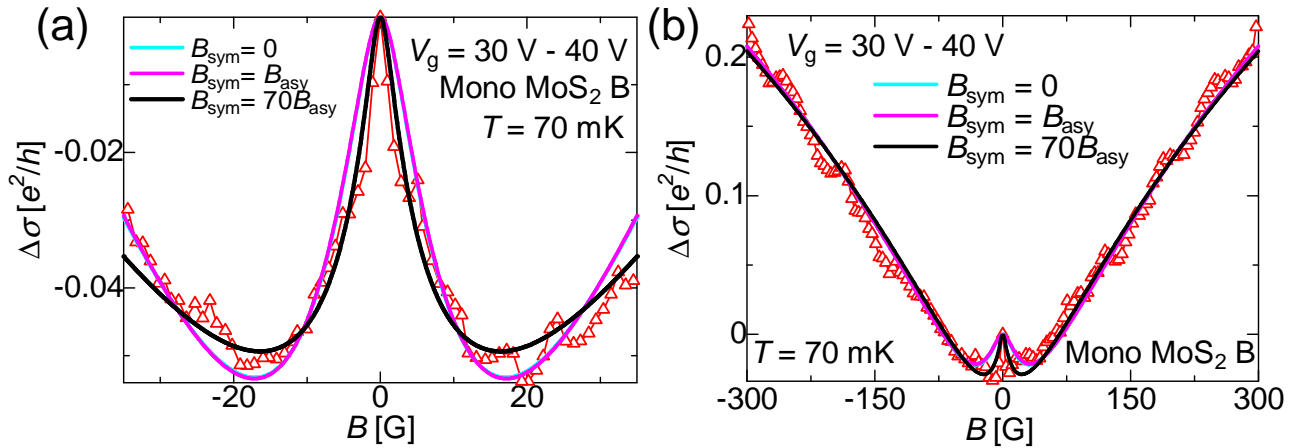}
\caption{(a) $\Delta \sigma (B)$ for $|B| <$ 35 G. The fits with no symmetric SOI ($B_{\rm sym}$ = 0) and the symmetric SOI equal to the asymmetric SOI ($B_{\rm sym}$ = $B_{\rm asy}$) deviate from the best fit ($B_{\rm sym}$ = 70$B_{\rm asy}$) especially close to the peak. Note that the two fits ($B_{\rm sym}$ = 0 and $B_{\rm sym}$ = $B_{\rm asy}$) almost overlap in the figure. (b) In high field region, there is no striking difference between the fits with different ratio of $B_{\rm sym}$ to $B_{\rm asy}$.}
\label{fig_MoS2}
\end{center}
\end{figure*}

\subsection{\label{sec:level2}Graphene/MoS$_2$ structures}
To explore the difference in the induced SOI on graphene from different TMDs, we also investigated SOI of graphene in proximity to MoS$_2$. MoS$_2$ has intrinsic SOI smaller than WSe$_2$ and WS$_2$. The calculated intrinsic spin-orbit splittings for the valence bands are 150 meV, 430 meV and 460 meV for MoS$_2$, WS$_2$ and WSe$_2$, respectively \cite{xiao}. Therefore if the SOI induced in graphene is provided by interface interactions with TMD, MoS$_2$ should induce smaller SOI in graphene than graphene/WSe$_2$ (or WS$_2$) structures. Based on this assumption we conducted low temperature magnetoresistance measurements on graphene on monolayer MoS$_2$ (Mono MoS$_2$) samples. Figure 4(c) displays a $\Delta \sigma (B)$ curve for one of the Mono MoS$_2$ samples. The WAL peak is observed around $B$ = 0 as for other graphene/TMD structures. However, in contrast to Mono WSe$_2$ and WS$_2$, $\Delta \sigma (B)$ strongly increases for higher $B$ region, and similar characteristics are observed for other gate voltage ranges. This reveals that the SOI induced in graphene by monolayer MoS$_2$ is smaller than that for Mono WSe$_2$ and WS$_2$ samples, and the amplitudes are similar to those of Bulk WSe$_2$ and WS$_2$. Another sample of graphene/monolayer MoS$_2$ with larger graphene mobility also yields comparably small induced SOI. In Fig. \ref{fig_MoS2} we show $\Delta \sigma (B)$ for Mono MoS$_2$ B in small field region (Fig. \ref{fig_MoS2}(a)) and in high field region (Fig. \ref{fig_MoS2}(b)). For the graphene/MoS$_2$ devices, it is essential to evaluate the amplitudes of SOI in small field region as discussed below.

\section{\label{sec:level1}Analysis\protect\\ }
\subsection{\label{sec:level2}Subtraction of the background signals}
\begin{figure*}[tb]
\begin{center}
\includegraphics[width=17cm,clip]{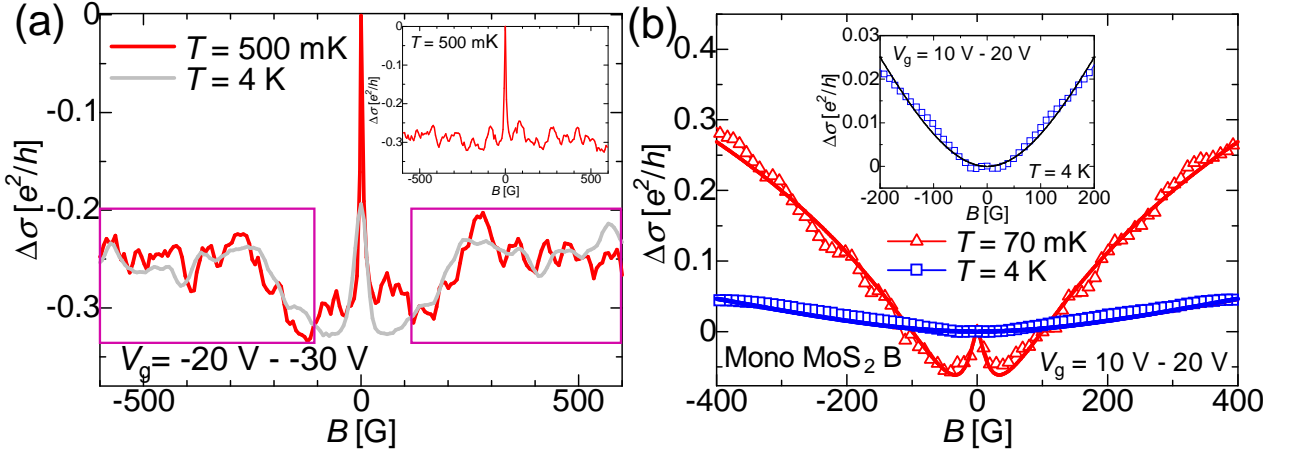}
\caption{\textbf{Removal of temperature independent components:}
(a) The original data at 500 mK and that at 4 K from Mono WS$_2$ A. In the inset we show $\Delta \sigma (B)$ obtained from Mono WS$_2$ A after the subtraction of the background signal. The subtraction procedure is written in the main text, and the subtracted regions are marked by the rectangles. (b) Comparison of the $\Delta \sigma (B)$ curves at 70 mK and 4 K for Mono MoS$_2$ B. It is clear that even for higher $B$ region, the shape of the two curves is drastically different at the two temperatures, indicating that there does not exist any temperature independent components. Theoretical fits are shown by the solid lines. Inset: The same curve of $\Delta \sigma (B)$ at 4 K as in the main figure with the fit based only on the weak localization term in (1) (namely, $\tau_{\rm asy} \rightarrow \infty$ and $\tau_{\rm so} \rightarrow \infty$). The upturn is well reproduced therefore it can be attributed to the weak localization contribution.}
\label{fig5}
\end{center}
\end{figure*}
As discussed in previous studies \cite{wang2, zihlmann, wakamura}, WAL signals are sometimes superimposed a top of temperature-independent magnetoresistance  backgrounds particularly for high $B$ regions \cite{wang2, zihlmann} which need to be subtracted for a proper analysis of WL and WAL signals. %For example, for the experimental data shown in the previous section, while such   background signals are subtracted for the Mono WS$_2$ curves with $V_g$ between $-$50 V and $-$60 V, no backgrounds are subtracted out side of this interval of gate voltage.
In previous studies \cite{zihlmann} this background was sometimes attributed to  a classical magnetoresistance contribution proportional to $B^2$, but this temperature-independent component in our case has a different shape.  Although the origin of these signals is still unclear, the temperature independence indicates that they stem from classical contributions rather than quantum contributions. Since the theoretical formula used to fit the experimental results considers only quantum contributions, it is justified to subtract these temperature independent components from the original signal. We note that this subtraction dramatically changes the estimation of the induced SOI 
%which is quite sensitive to the high field regions is important to determine $\tau_{\rm so}$,
 as pointed out in previous reports \cite{wang2, roche2, grbic}. For Mono WS$_2$, a temperature independent part is observed particularly for $V_g$ $<$ 0. By contrast, the  existence of a temperature independent magnetoconductance background in Mono WSe$_2 $ only appears for some gate voltage ranges in both electron-doped and hole-doped regions. 

In Fig. \ref{fig5}(a) we show the original data of $\Delta \sigma (B)$ from Mono WS$_2$ taken at 500 mK and at 4 K. $\Delta \sigma (B)$ at 500 mK after the subtraction of the temperature independent magnetoresistive component is displayed in the inset. Note that $\Delta \sigma (B)$ at 4 K is shifted vertically so that the temperature independent part overlaps with that at 500 mK. One can clearly see that for $|B|$ $>$ 100 G, $\Delta \sigma (B)$ is the same at 500 mK and 4 K except for the strong conductance fluctuations observed at 500 mK. Keeping the original data points for $|B|$ $<$ 100 G, we subtract the shifted $\Delta \sigma (B)$ at 4 K from the original $\Delta \sigma (B)$ at 500 mK, namely, $\Delta \sigma$ ($B$, 500 mK, subtracted) = $\Delta \sigma$ ($B$, 500 mK, original) - $\Delta \sigma$ ($B$, 4 K, shifted).  After subtraction, the $\Delta \sigma (B)$ curve is flat for high field regions. The subtraction of background signals is performed for Mono WSe$_2$ and WS$_2$ for certain ranges of gate voltage , but not for the other samples. For example, Mono MoS$_2$ exhibits a temperature dependent upturn, as shown in Fig. \ref{fig5}(b) which is included in the analysis of the quantum conductivity correction.
\subsection{\label{sec:level2}Analysis of the weak antilocalization signals to evaluate SOI}    
\begin{figure}[b]
\includegraphics[width=8cm,clip]{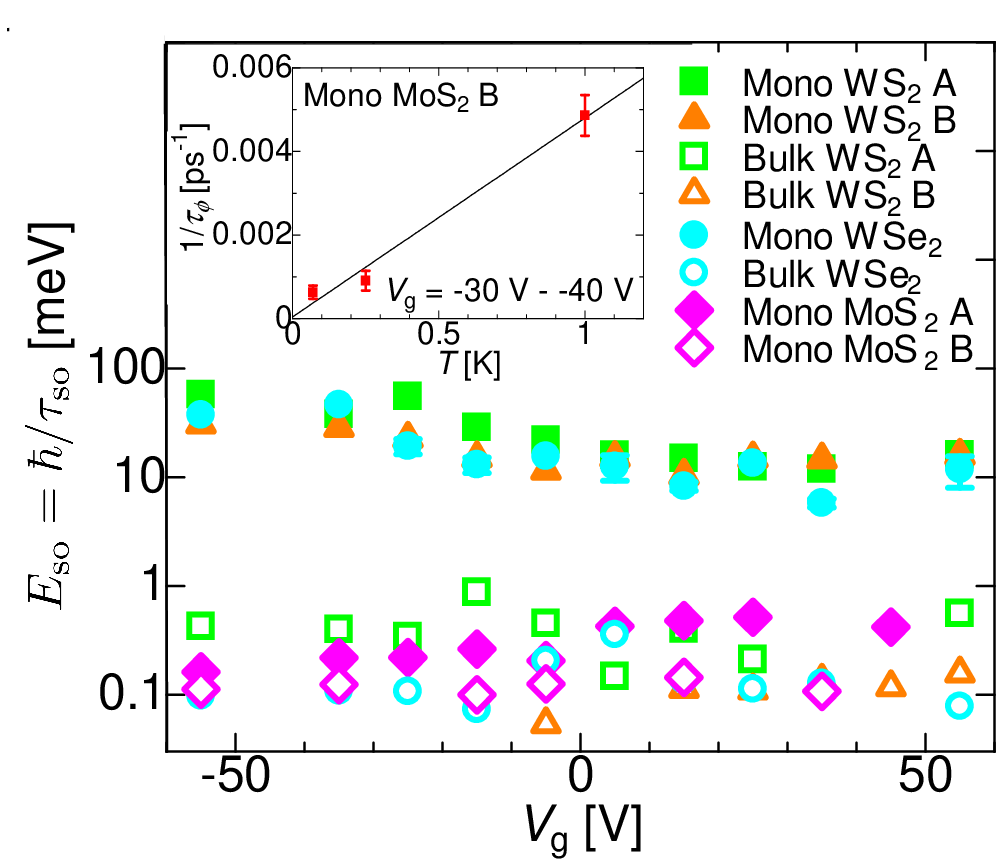}
\caption{Comparison of the spin-orbit energy ($E_{\rm so}$) estimated from the theoretical fits for each graphene/TMD heterostructure. The eight samples can be categorized into the two groups, the group with $E_{\rm so} \gtrsim$ 10 mV (Mono WSe$_2$ and WS$_2$) and the one with $E_{\rm so} \lesssim$ 1 meV (Mono MoS$_2$, Bulk WSe$_2$ and WS$_2$). In the inset we show the temperature dependence of $\tau_\phi$ for the sample Mono MoS$_2$ B as an example. The experimental data is consistent with the relation $\tau_\phi^{-1} \propto T$.}
\label{fig7}
\end{figure}

Based on the above considerations, we attempt to fit the experimental results using the equation (\ref{eq1}) for the weak antilocalization. We first discuss the total amplitudes of the SOI determined by $\tau_{\rm so}$. In Fig. \ref{fig7} we plot the spin-orbit scattering strength defined as $E_{\rm so} = \hbar / \tau_{\rm so}$ for each system as a function of the gate voltage. These systems can be divided into two groups, one that exhibits strong spin-orbit scattering ($E_{\rm so} \sim $ 10 meV) and the other that shows 
moderate spin-orbit scattering ($E_{\rm so} \lesssim$ 1 meV). Clearly, Mono WSe$_2$ and the two Mono WS$_2$ samples belong to the former group and yield strong $E_{\rm so}$ that amounts to 10 meV or even larger. In contrast, $E_{\rm so}$ for Bulk WSe$_2$ and Bulk WS$_2$ are in the latter group and $E_{\rm so}$ is more than an order of magnitude smaller. This striking difference between graphene/monolayer TMD and graphene/bulk TMD is consistent with our previous study \cite{wakamura}. On the other hand, the two Mono MoS$_2$ samples exhibit $E_{\rm so}$ $<$ 1 meV, similar to Bulk WSe$_2$ and Bulk WS$_2$. Therefore the amplitudes of the induced SOI depends not only on the thickness of the TMD layer but also on the type of TMD. As briefly discussed in the previous section, given that the intrinsic SOI of monolayer MoS$_2$ is three times smaller than that of monolayer WSe$_2$ and WS$_2$, it is reasonable that the induced SOI in graphene is smaller for the graphene/MoS$_2$ system. We note that for the samples with strong SOI (Mono WS$_2$ and Mono WSe$_2$), the estimated $\tau_{\rm so}$ is close to the momentum relaxation time $\tau_p$ and for some gate voltage ranges it is even smaller ($\tau_{\rm so} \lesssim \tau_p$). Since this limit is out of the validity range of formula (\ref{eq1}), in the discussion (on the spin relaxation mechanism) below we exclude the data points in this limit.

\subsection{\label{sec:level2}Effect of intervalley scattering}
\begin{figure*}
\begin{center}
\includegraphics[width=17cm,clip]{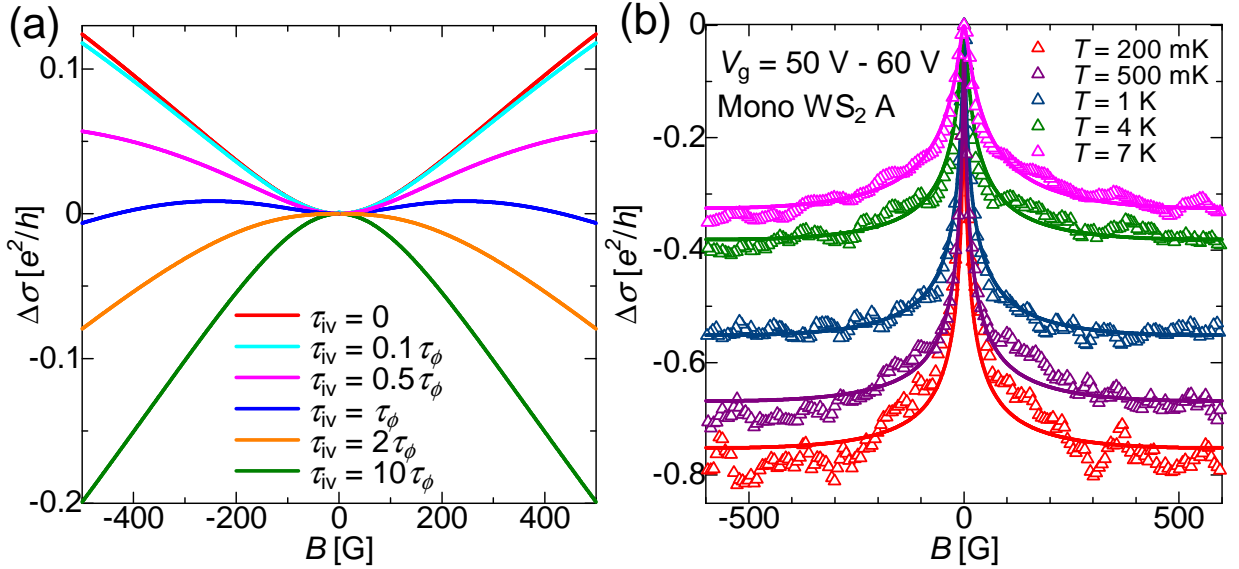}
\caption{
(a) Simulated curves of $\Delta \sigma (B)$ based on equation (\ref{eq_iv}) with $B_\phi = \hbar/4eD\tau_\phi$=0.01 with different ratio of $\tau_\phi$/$\tau_{iv}$. The shape of $\Delta \sigma (B)$ dramatically changes for $\tau_{iv} \gtrsim \tau_\phi$. (b) Experimental data of $\Delta \sigma (B)$ from Mono WS$_2$ A. The flat tails in high field region are observed over a broad range of temperature, indicating that the system is in the limit $\tau_{iv} \ll \tau_\phi$. The solid lines are the fits based on (1).}
\label{fig_iv}
\end{center}
\end{figure*}
For all of the fits we have assumed that $\tau_{iv} \sim \tau_p \ll \tau_\phi$. By doing so, we extract parameters which are consistent for all samples investigated in a wide range of gate voltages and temperatures. On the other hand, a recent theoretical study has proposed that the WAL peaks observed in our experiments without upturn in high fields could be reproduced in the limit $\tau_\phi \sim \tau_{\rm so} \ll \tau_{iv}$ \cite{ilic}. In this limit weak antilocalization can be driven by the chirality \cite{Tikhonenko}, if the induced SOI is small. To clarify the effect of intervalley scatterings on $\Delta \sigma (B)$ in high field region, we plot calculated curves where SOI is small and the ratio between $\tau_\phi$ and $\tau_{iv}$ determines the shape of $\Delta \sigma (B)$. In this case $\Delta \sigma (B)$ can be expressed as \cite{mccann2}
\begin{multline}
\Delta \sigma (B) = \frac{e^2}{\pi h} \left[F \left( \frac{\tau_B^{-1}}{\tau_\phi^{-1}} \right) - F \left( \frac{\tau_B^{-1}}{\tau_\phi^{-1} + 2 \tau_{iv}^{-1}} \right) \right.\\
\left. -2 F \left( \frac{\tau_B^{-1}}{\tau_\phi^{-1} + \tau_{iv}^{-1}} \right) \right],
\label{eq_iv}
\end{multline}
where we neglect the contribution from intravalley scatterings for simplicity. In Fig. \ref{fig_iv}(a) we show the simulated curves with different ratio between $\tau_\phi$ and $\tau_{iv}$. In the limit of $\tau_{iv} \ll \tau_\phi$, the shape of $\Delta \sigma (B)$ is insensitive to the change of the ratio $\tau_\phi$/$\tau_{iv}$. However, when $\tau_{iv} \gtrsim \tau_\phi$, $\Delta \sigma (B)$ exhibits a crossover from upturn to downturn feature, and is highly sensitive to the change of $\tau_\phi$/$\tau_{iv}$. It is important to note that our samples with the strong induced SOI (Mono WSe$_2$ and Mono WS$_2$) show flat tails in the high field region over a broad range of temperature (between 250 mK to 7 K, see Fig. \ref{fig_iv}(b) as an example). Because $\tau_{iv}$ is independent of tempearture and $\tau_\phi$ varies dramatically in this range of temperature (see the inset of Fig. \ref{fig7} and Fig. \ref{fig9}), the ratio $\tau_\phi$/$\tau_{iv}$ also changes a lot. Therefore the flat tails observed in the broad range of temperature can only be explained in the limit $\tau_{iv} \ll \tau_\phi$. 

%However, such an assumption is not consistent with the observation of a clear weak localization behavior after the peak at $B=0$ for Bulk WSe$_2$, Bulk WS$_2$ and Mono MoS$_2$ samples for which SOI interactions are weak.
%In the inset of Fig. \ref{fig5}(b) we show the fit of the upturn of Mono MoS$_2$ only by the weak localization contribution (1st term) of the equation (1). Good agreement between the experimental data and the fit demonstrate that the upturn derives from the weak localization contribution.%We comment on recently proposed another formula for weak antilocalization in graphene that includes the VZ contributions. Theoretically 
\subsection{\label{sec:level2}Symmetry of the induced SOI}
\begin{figure}[tb!]
\begin{center}
\includegraphics[width=8cm,clip]{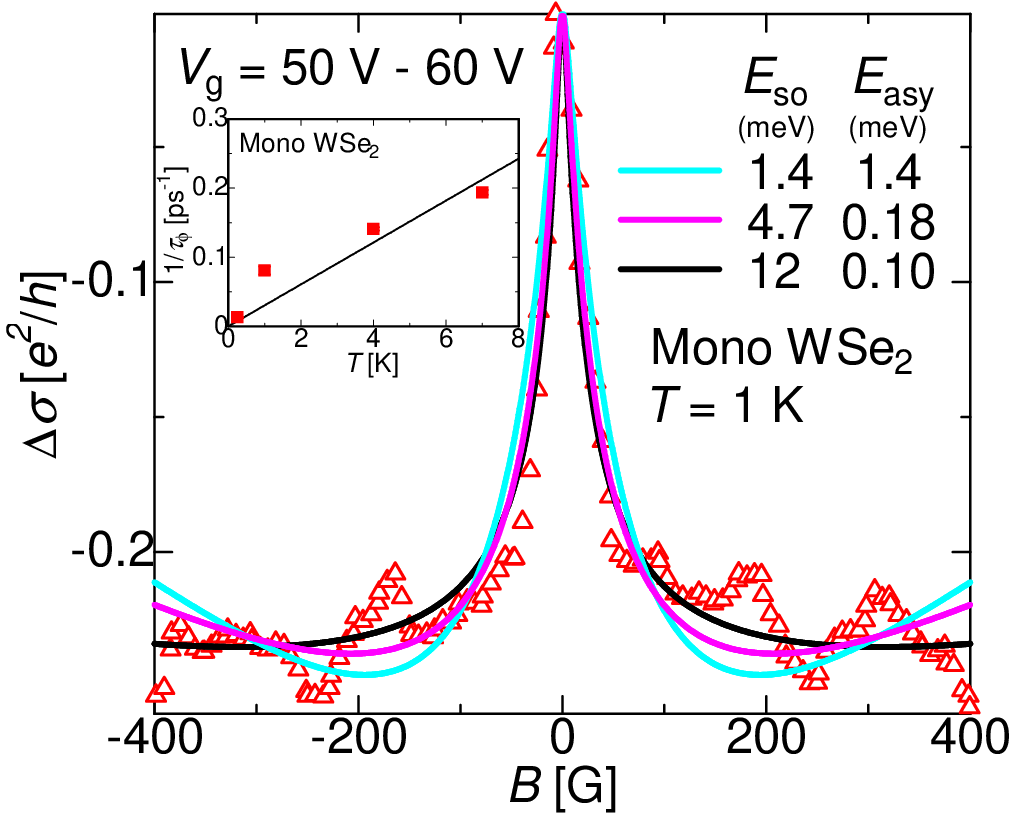}
\caption{\textbf{Sensitivity of the fits to the ratio $\tau_{\rm sym}/\tau_{\rm asy}$:} Different fits are shown for different values of $\tau_{\rm sym}/\tau_{\rm asy}$: When there is no symmetric contribution (light blue curve), the fit strongly deviates from the experimental points. In contrast, when $\tau_{\rm sym} \ll \tau_{\rm asy}$ and $E_{\rm so} \gtrsim$ 12 meV, we can well reproduce the experimental data. Inset: Temperature dependence of 1/$\tau_\phi$ for Mono WSe$_2$ for $V_g$ between 50 V and 60 V.}
\label{fig9}
\end{center}
\end{figure}

As explained above, equation (\ref{eq1}) includes two fitting parameters $\tau_{\rm so}^{-1} = \tau_{\rm sym}^{-1} + \tau_{\rm asy}^{-1}$ and $\tau_{\rm asy}$. KM and VZ SOI, which are $z \rightarrow -z$ symmetric, determine $\tau_{\rm sym}$ whereas $\tau_{\rm asy}$ is attributed to the $z \rightarrow -z$ asymmetric Rashba SOI. Identifying the symmetry of the induced SOI is particularly important to provoke intriguing phenomena such as the QSH effect in graphene. The seminal paper by Kane and Mele \cite{kane1, kane2} reported that  a dominant $z \rightarrow -z$ symmetric SOI is required for the QSH state. From the fitting based on (\ref{eq1}), we can determine the ratio between the symmetric and asymmetric contributions to the induced SOI because both $\tau_{\rm so}$($\tau_{\rm sym}$) and $\tau_{\rm asy}$ are fitting parameters. In our previous paper \cite{wakamura} we demonstrated that for WS$_2$ systems the symmetric SOI is dominant. To evaluate the symmetry of the induced SOI in other systems, in Fig. \ref{fig9} we show the fits based on different ratios between $\tau_{\rm sym}$ and $\tau_{\rm asy}$ in (\ref{eq1}). As seen in Fig. \ref{fig9}, when there is no symmetric contribution, the fitting curve (shown in light blue) exhibits a clear upturn in high $B$ and largely deviates from the experimental data in the smaller $B$ region as well. We note that for $\tau_{\rm sym} \sim \tau_{\rm asy}$  it is not possible to fit correctly the data either. To reproduce the flat tail of the experimental data in the high $B$ region a dominant symmetric contribution is required, and the best fit is obtained when the spin-orbit scattering strength ($E_{\rm so}$) is $E_{\rm so} \gtrsim$ 12 meV. This dominant symmetric contribution is also consistent with the work of \cite{zihlmann}.

A similar analysis has been performed for all investigated samples. We note that for $\Delta \sigma (B)$ with a temperature-dependent upturn at high $B$ as shown in Fig. \ref{fig5}(b), it is essential to carry out the analysis in small field region. This is because the upturn is attributed to the weak localization contribution, irrelevant to SOI. In the inset of Fig. \ref{fig5}(b), we display the fit of $\Delta \sigma (B)$ up to $B$ = 200 G only by taking into account the weak localization limit ($\tau_{\rm asy} \rightarrow \infty$ and $\tau_{\rm so} \rightarrow \infty$) in (\ref{eq1}). The fit reproduces the experimental data well, indicating that weak localization is the main contribution in this limit. On the other hand, it also implies that in this field range there is a large ambiguity to determine $\tau_{\rm asy}$ and $\tau_{\rm so}$ precisely. In Fig. \ref{fig_MoS2}(a) and (b) we show the fits of $\Delta \sigma (B)$ from MoS$_2$ B in small field (Fig. \ref{fig_MoS2}(a)) and large field (Fig. \ref{fig_MoS2}(b)) region with different ratio between $B_{\rm sym}$ and $B_{\rm asy}$, where $B_{\rm X} = \hbar/4eD\tau_{\rm X}$, X = asy or so. While the difference is not as striking as in the case of Mono WSe$_2$ and Mono WS$_2$, in the small field region the large symmetric SOI ($B_{\rm sym}$ = 70$B_{\rm asy}$) provides the best fit in comparison with the other two curves ($B_{\rm sym}$ = 0 and $B_{\rm sym}$ = $B_{\rm asy}$) as examples. By contrast, in the high field region, the fits mainly account for a large number of points in the upturn part, where weak localization plays a major role. Indeed, the three fits with different ratio between $B_{\rm sym}$ and $B_{\rm asy}$ provide almost similar curves. Thus to determine the spin-orbit parameters ($B_{\rm asy}$ and $B_{\rm so}$) accurately, it is indispensable to carry out analysis in a small field region for the samples with temperature-dependent upturn.
% While $\sqrt{\tau_{\rm asy}/\tau_{\rm sym}}$ considerably varies depending on the sample and gate voltage range, 

In Table \ref{tab:table2} we provide the square root of the ratio between $\tau_{\rm asy}$ and $\tau_{\rm sym}$ ($\sqrt{\tau_{\rm asy}/\tau_{\rm sym}}$) for all investigated samples. These large ratios of $\sqrt{\tau_{\rm asy}/\tau_{\rm sym}}$ are consistent with other studies \cite{wang2, valenzuera, vanwees, zihlmann}.
% A dominant contribution  of $z \rightarrow -z$ symmetric SOI is  clearly found for all samples.

%\begin{figure}
%\includegraphics[width=8cm,clip]{tau_asy.eps}
%\caption{The fits by modulating the symmetric part of SOI ($\tau_{\rm sym}$). When there is no symmetric contribution (light blue curve), the fit strongly deviates from the experimental points. In contrast, when $\tau_{\rm sym} \ll \tau_{\rm asy}$ and $E_{\rm so} \gtrsim$ 12 meV, we can well reproduce the experimental data.}
%\label{tau_asy}
%\end{figure}
%In Fig. \ref{tau_asy} we show $\tau_{\rm asy}$ for different samples as a function of $V_g$. $\tau_{\rm asy}$ ranges from 1 to 100 ps depending on the sample, but all of them are much larger than $\tau_{\rm sym}$ 
\begin{table}[b]%The best place to locate the table environment is directly after its first reference in text
\caption{\label{tab:table2} Minumum and maximum values of the square root of the $\tau_{\rm asy}/\tau_{\rm sym}$ ratio giving equally acceptable fits of the data. A large difference between $\tau_{\rm asy}$ and $\tau_{\rm sym}$ is observed for all samples.
}
\begin{ruledtabular}
\begin{tabular}{lc}
\textrm{Sample}&
\textrm{$\sqrt{\tau_{\rm asy}/\tau_{\rm sym}}$}\\
\colrule
Mono WSe$_2$ & 5.2 - 16 \\
Mono WS$_2$ A & 25 - 53 \\
Mono WS$_2$ B & 22 - 56 \\
Mono MoS$_2$ A & 16 - 64 \\
Mono MoS$_2$ B & 2.7 - 13 \\
Bulk WSe$_2$ & 3.5 - 10 \\
Bulk WS$_2$ A & 2.6 - 21 \\
Bulk WS$_2$ B & 1.9 - 15 \\
\end{tabular}
\end{ruledtabular}
\end{table}
\subsection{\label{sec:level2}Identification of the dominant SOI type from the spin relaxation mechanism}
Now that we have found that the $z \rightarrow -z$ symmetric SOI is dominant, we next determine the dominant spin relaxation mechanism since this symmetric SOI is composed of two contributions, KM and VZ SOI.  For graphene, there are two possible spin relaxation mechanisms: Elliot-Yafet (EY) and D'yakonov-Perel (DP) mechanism \cite{zutic, fabian, huetas, ochoa2}. KM SOI contributes to the EY mechanism \cite{mccann} since the DP mechanism requires spin-splitting due to inversion symmetry breaking. On the other hand, since VZ SOI \cite{roche} arises from broken sublattice symmetry, it causes DP-type spin relaxation. We here neglect the contributions from Rashba SOI because the estimates for $\tau_{\rm asy}$ indicate that Rashba SOI is smaller than the other types of SOI. As reported in previous studies \cite{wakamura, zomer}, each contribution (EY or DP) can be determined by fitting the relation between $\tau_{\rm so}$ and the momentum relaxation time $\tau_p$ following the equation
\begin{equation}
\frac{\varepsilon_F^2 \tau_p}{\tau_{\rm so}} = \Delta_{\rm EY}^2 + \left( \frac{4 \Delta_{\rm DP}^2}{\hbar^2} \right)
\varepsilon_F^2 \tau_p^2
\label{eq_spin_relax}
\end{equation}
where $\Delta_{\rm EY(DP)}$ is the amplitude of spin-orbit coupling leading to EY (DP) mechanism and $\varepsilon_F$ is the Fermi energy.
\begin{figure*}
\begin{center}
\includegraphics[width=17cm,clip]{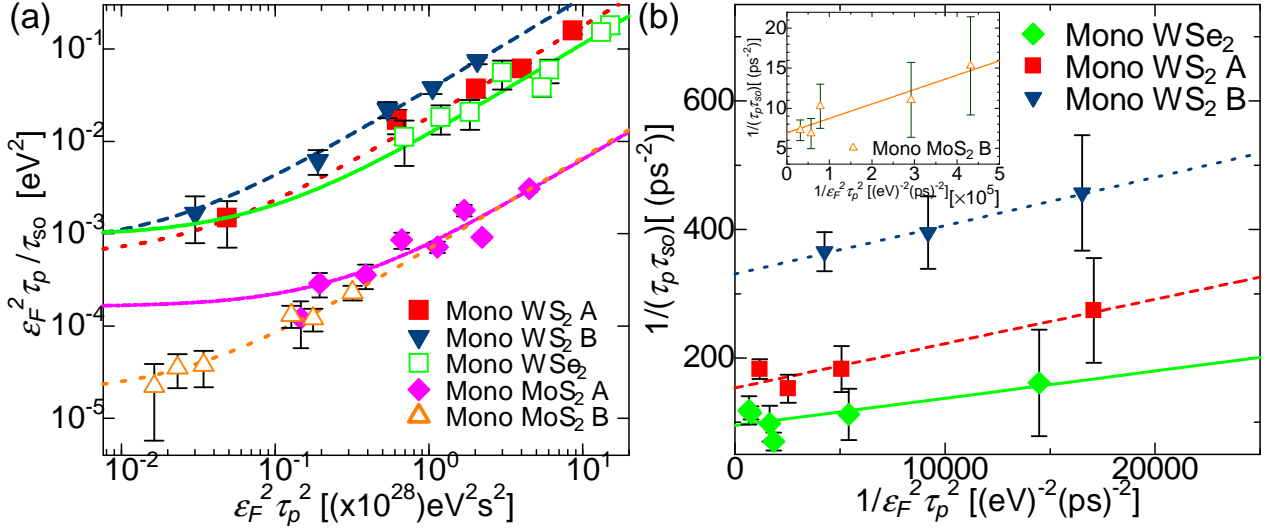}
\caption{\textbf{Extraction of $\Delta_{\rm EY}$ and $\Delta_{\rm DP}$ to identify the dominant SOI for each heterostructure:}
(a) Log-scale plot of $\varepsilon_F^2 \tau_p / \tau_{\rm so}$ as a function of $\varepsilon_F^2 \tau_p^2$. The non-zero $y$-axis intercept and the deviation from the linearity are due to the EY-type spin relaxation. The error bars are obtained by considering the standard deviation of the Fermi energy ($\delta E_F$) due to electron-hole puddles. (b) Using the same experimental data as (a) we plot them in a different way based on (\ref{eq_spin_relax2}) to clarify the EY contribution. The positive slope is clearly seen as a result of the EY contribution for the three samples. The inset shows the same plot for Mono MoS$_2$ B.}
\label{fig8}
\end{center}
\end{figure*}
In Fig. \ref{fig8}(a) we plot relation (\ref{eq_spin_relax}) for samples with monolayer TMDs in log scale. For all samples including the ones not shown in the figure we found that $\Delta_{\rm EY}$ is larger than $\Delta_{\rm DP}$, and the ratio $\Delta_{\rm EY}/\Delta_{\rm DP}$ varies from 5 to 31 depending on the sample. In the figure of the fitting based on (\ref{eq_spin_relax}), the EY contribution leads to a nonzero $y$-axis intercept, and to the deviations from the straight line in Fig. \ref{fig8}(a). Another way of visualizing the EY contribution is obtained by dividing both sides of (\ref{eq_spin_relax}) by $\varepsilon_F^2 \tau_p^2$, leading to
\begin{equation}
\frac{1}{\tau_p \tau_{\rm so}} = \Delta_{\rm EY}^2 \left( \frac{1}{\varepsilon_F^2 \tau_p^2} \right) + \frac{4 \Delta_{\rm DP}^2}{\hbar^2}.
\label{eq_spin_relax2}
\end{equation}
Figure \ref{fig8}(b) shows the relation (\ref{eq_spin_relax2}) for three samples with the strongest SOI (Mono WSe$_2$, Mono WS$_2$ A and B). We also plot this relation for Mono MoS$_2$ B in the inset of Fig. \ref{fig8}(b). In these plots the slope of the fit corresponds to the EY contribution. The positive slope that represents the contribution from the EY mechanism is clear for each sample, demonstrating the existence of KM SOI. The fits in Fig. \ref{fig8}(a) and (b) provide somewhat different $\Delta_{\rm EY}$ and $\Delta_{\rm DP}$, therefore in Table \ref{tab:table1} we show the averaged values over the two fits.

The most important difference between the EY and DP spin relaxation mechanism is the different dependence of $\tau_{\rm so}$ on the momentum scattering time $\tau_p$. While $\tau_{\rm so} \propto \tau_p$ for the EY mechanism, for the DP $\tau_{\rm so} \propto \tau_p^{-1}$. In the case of graphene since $\tau_p$ can be modulated by $V_g$ the ratio between the EY and DP contribution also depends on $V_g$. This means that depending on $V_g$ the contributions from each SOI (KM or VZ SOI) vary. Hence the EY mechanism, or KM SOI plays an important role particularly around the Dirac point.

We note that if we assume that VZ SOI is the only source of DP spin relaxation, we can replace $\tau_p$ with $\tau_{iv}$ in (\ref{eq_spin_relax}), where $\tau_{iv}$ is the intervalley scattering time \cite{roche}. Since $\tau_{iv}$ is always larger than $\tau_p$, $\Delta_{\rm DP}$ becomes then even smaller. 
\begin{table}[b]%The best place to locate the table environment is directly after its first reference in text
\caption{\label{tab:table1} $\Delta_{\rm EY}$ and $\Delta_{\rm DP}$ obtained for each sample by fitting the experimental results using (\ref{eq_spin_relax}) and (\ref{eq_spin_relax2}) separately and averaged over the two values.}
\begin{ruledtabular}
\begin{tabular}{lcc}
\textrm{Sample}&
\textrm{$\Delta_{\rm EY}$ [meV]}&
%\multicolumn{1}{c}{\textrm{$\Delta_{\rm DP}$}}&
\textrm{$\Delta_{\rm DP}$ [meV]}\\
\colrule
Mono WSe$_2$ & 48.0 $\pm$ 12.2 & 3.3 $\pm$ 0.10\\
Mono WS$_2$ A & 27.4 $\pm$ 2.8 & 4.4 $\pm$ 0.047\\
Mono WS$_2$ B & 32.9 $\pm$ 3.8 & 6.4 $\pm$ 0.14\\
Mono MoS$_2$ A & 10.3 $\pm$ 1.7 & 0.87 $\pm$ 0.035\\
Mono MoS$_2$ B & 4.3 $\pm$ 0.035 & 0.86 $\pm$ 3.5$\times 10^{-3}$\\
Bulk WSe$_2$ & 11.6 $\pm$ 2.1 & 0.38 $\pm$ 0.013\\
Bulk WS$_2$ A & 8.9 $\pm$ 3.7 & 0.73 $\pm$ 0.028\\
Bulk WS$_2$ B & - & 0.72 $\pm$ 0.022\\
\end{tabular}
\end{ruledtabular}
\end{table}%
\section{\label{sec:level1}Discussions\protect\\ }
\subsection{\label{sec:level2}Possibility of VZ SOI}
In the previous section we pointed out the possibility to induce KM SOI in graphene. In contrast, recent \textit{ab initio} calculations propose the scenario of VZ SOI as the dominant part of the induced SOI. The previous study by Frank \textit{et al}. \cite{frank} revealed that VZ SOI generates edge states as KM SOI, but they are not topologically protected. Thus it is of great importance to discuss the possibility of inducing VZ SOI to determine if one can realize the QSH state in graphene on TMD. VZ SOI originates from the $A-B$ sublattice symmetry breaking in graphene. If the sublattice symmetry is broken, different values of the spin-orbit parameter $\lambda_{\rm I}^{\rm A}$  and $\lambda_{\rm I}^{\rm B}$ can appear in the Hamiltonian for symmetric SOI ($H_{\rm sym}$) depending on the sublattice. Thus $H_{\rm sym}$ can be written as \cite{gmitra3}:
\begin{equation}
H_{\rm sym}= \frac{1}{2} [\lambda_{\rm I}^{\rm A} (\sigma_z + \sigma_0) + \lambda_{\rm I}^{\rm B} (\sigma_z - \sigma_0)] \tau_z s_z,
\label{eq3}
\end{equation}
where $\sigma_z$, $\tau_z$ and $s_z$ denote Pauli matrix for sublattice spin, valley spin and real spin, respectively. $\sigma_0$ is the unit matrix in the sublattice space. The terms in equation (\ref{eq3}) can then be categorized into two groups according to their dependence on sublattice spin. The first one is proportional to $\sigma_z$, as the original KM SOI, and expressed as:
\begin{equation}
H_{\rm KM}= \lambda_{\rm KM} \sigma_z \tau_z s_z
\label{eq4}
\end{equation}
where $\lambda_{\rm KM} = (\lambda_{\rm I}^{\rm A} + \lambda_{\rm I}^{\rm B})/2$. %In (\ref{eq4}) we explicitly write "KM" instead of "I" to distinguish it from the original intrinsic SOI written in (\ref{eq3}).
The second group reads
\begin{equation}
H_{\rm VZ}= \lambda_{\rm VZ} \sigma_0 \tau_z s_z
\label{eq5}
\end{equation}
where $\lambda_{\rm VZ} = (\lambda_{\rm I}^{\rm A} - \lambda_{\rm I}^{\rm B})/2$. This term is called VZ SOI. To obtain nonzero VZ SOI, $\lambda_{\rm I}^{\rm A} \neq \lambda_{\rm I}^{\rm B}$ is required, thus breaking graphene sublattice symmetry is indispensable.

In the \textit{ab initio} calculations, the unit cell is composed of (e.g.) 5$\times$5 graphene and 4$\times$4 TMD supercells whose bond length have been relaxed in order to generate a perfectly periodic lattice. Experimentally, however, due to a large lattice mismatch between graphene and TMDs ($\sim$28 \%) such a perfect periodicity does not exist \cite{gmitra3}. Absence of perfect periodicity is also confirmed by the observations of pseudoperiodic faint Moir\'e pattern \cite{moire1, moire2}. Therefore graphene/TMD superlattices are never perfectly periodic \cite{moire3} unless the relative rotational angle between two lattices (graphene and TMD) is carefully selected. Slight deviations from the perfect periodicity on a small scale with a few lattices do not provide a considerable effect, but may result in a large deviation in a wider scale and give rise to an equal spin-orbit potential for sublattices A and B on average. Therefore in real graphene/TMD heterostructures the sublattice symmetry may be locally broken on a small scale, but when the average is taken over the mean free path (or intervalley scattering length), which sets a length scale for VZ SOI, the effect of sublattice symmetry breaking might vanish or be considerably suppressed.

On the other hand, the analysis of the spin relaxation mechanisms demonstrated a clear DP contribution although it is much smaller than the EY contribution. Since the other types of SOI (Rashba and pseudospin inversion asymmetry (PIA) SOI) that can provide the DP mechanism are $z \rightarrow -z$ asymmetric, and found to be small by the WAL measurements, we cannot rule out the contribution from VZ SOI to the observed $z \rightarrow -z$ symmetric SOI. Further experimental and theoretical investigations are required for this issue.

It is also interesting to discuss the strength of sublattice symmetry breaking in graphene caused by the underlying TMD layer. A previous report on giant Rashba splitting in graphene due to hybridization with gold \cite{gold} provides important information. In the calculated band structures, strong Rashba splitting was found with gold adatoms on top of a given sublattice (e.g. sublattice $A$). It is clear that these gold adatoms break sublattice symmetry, but no sublattice gap is opened between valence and conduction bands unless the distance between graphene and gold atom is unrealistically closer ($\sim$2.4 \AA) than the equilibrium length ($\sim$3.4 \AA). In the case of graphene on TMD, graphene's $p_z$ orbitals couple to the $d$ orbitals of transition metals or $p$ orbitals of chalcogen. Taking into account the distance between graphene and TMD ($\sim$3.4 \AA) and also the distance between a transition metal and a chalcogen layer ($\sim$1.7 \AA), it is possible that the sublattice symmetry breaking effect may be small. Angular resolved photoemission spectroscopy (ARPES) studies reported an intact bandstructures of graphene close to the Dirac point in graphene/MoS$_2$ \cite{arpes1, arpes2} and graphene/WS$_2$ \cite{arpes3} heterostructures. It is found that only when the relative angle between graphene and TMD lattice is carefully chosen, minigaps are obtained at high binding energies \cite{arpes2}. Based on these previous reports, the effects of the TMD underlayer on the graphene bandstructures may be weak.    

\subsection{\label{sec:level2}Suppressed Rashba SOI}
From the analysis on the WAL signals we concluded that the induced Rashba SOI is small in comparison with $z \rightarrow -z$ symmetric SOI. Naively, one would have expected instead that strong Rashba SOI is induced in graphene/TMD heterostructures due to the inversion symmetry breaking by the TMD layer. Indeed, some of the previous studies on inducing SOI in graphene by $d$-electron heavy adatoms revealed that strong Rashba SOI is induced \cite{gold, adatom5, adatom6}. One of the important differences between graphene/TMD systems and heavy adatoms (e.g. Au) on graphene is that charge transfer between graphene and TMD layers is much weaker than that between adatoms and graphene, as reported by recent ARPES measurements \cite{gold, arpes1, arpes2, arpes3}. As a result, only a small electric dipole is formed in the graphene/TMD systems compared with adatoms on graphene \cite{moire3}. Since a crucial role of charge transfer for Rashba SOI was pointed out theoretically \cite{abdel}, weak charge transfer effect may be the reason for the small Rashba SOI induced in graphene/TMD heterostructures. 

Suppressed Rashba and large SOI proportional to $s_z$ are also consistent with the previous measurements of nonlocal voltages generated by the (inverse) spin Hall effect (SHE) with strong SOI \cite{avsar2}. For the SHE, the relation between charge currents $\vec{J_c}$, spin polarization of charge currents $\vec{s}$ and generated spin currents $\vec{J_s}$ is expressed as $\vec{J_s} \propto \vec{s} \times \vec{J_c}$ \cite{takahashi}. To detect nonlocal voltages in a "H"-shaped device via the SHE and its inverse, $\vec{s}$ needs to be out-of-plane since $\vec{J_c}$ and $\vec{J_s}$ are both inplane. While Rashba SOI provides inplane spin polarization, for KM and VZ SOI the induced spin polarization is out-of-plane. Therefore dominant KM or VZ SOI are required to detect the nonlocal voltages demonstrated in \cite{avsar2}.
\subsection{\label{sec:level2}Difference in the amplitudes of the induced SOI in graphene among different TMDs}

In the previous sections we demonstrated that there are clear differences in the amplitudes of the SOI induced in graphene by different TMDs. We first discuss the difference among monolayer TMDs.

The analysis of  WAL signals showed that MoS$_2$ induces a spin-orbit scattering rate in graphene one order of magnitude smaller than WSe$_2$ and WS$_2$. It is known that MoS$_2$'s intrinsic SOI in the valence band is three times smaller than that of WSe$_2$ and WS$_2$ in the valence band \cite{xiao}. Considering the relation (\ref{eq_spin_relax}), the inverse spin-orbit time ($\tau_{\rm so}^{-1}$) is proportional to the square of the spin-orbit energy. Therefore the smaller SOI induced in graphene by MoS$_2$ is in agreement with the one order of magnitude difference expected in the spin-orbit amplitudes between graphene/WSe$_2$(or WS$_2$) and graphene/MoS$_2$ heterostructures.

On the other hand, recent ARPES measurements have revealed that the Dirac cone of graphene is closer to the conduction band edge than to the valence band edge of monolayer TMDs \cite{arpes1, arpes2, arpes3}. The spin-splitting of the conduction band edge of monolayer TMDs is neglected in the first approximation due to the $d_{z^2}$ nature of the orbital, which has zero orbital magnetic quantum number ($m_z=0$) \cite{xiao}. However, recent density functional theory (DFT) calculations point out the importance of the spin-splitting even for the conduction band \cite{kosmider, burkard} and provide different spin-splitting estimated for different TMDs. It is very likely that these values have also to be considered to compare the differents  SOI  induced in graphene. %These differences in intrinsic SOI of TMD may affect the observed differences in the SOI 

We also observed a clear difference in the amplitudes of the induced SOI in graphene between monolayer and bulk of the same TMDs. This difference may arise from different surface matching. In general graphene on TMDs is not perfectly flat and there are bubbles and ripples \cite{wang2}. Monolayer TMDs are more flexible than bulk TMDs so they could better follow the curvature of the graphene layer. The net area where graphene covered by TMD would thus be greater and the total SOI induced in graphene may be enhanced.

In our previous paper we also proposed that the band structure differences between monolayer and bulk TMD may affect the amplitudes of the induced SOI. However, it is not so likely because no striking differences in the band structure are observed in ARPES measurements between graphene/monolayer MoS$_2$ and graphene/bulk MoS$_2$ samples \cite{arpes1, arpes2}.
\section{\label{sec:level1}Conclusions\protect\\ }
In conclusion, we successfully induced strong SOI in graphene by exploiting heterostructures with different TMDs of different thickness. By comparing each system, we observed both universal and nonuniversal characters. Monolayer tungsten-based TMDs (WSe$_2$ and WS$_2$) induce stronger SOI in graphene, while monolayer MoS$_2$ induces SOI one order of magnitude smaller. For WSe$_2$ and WS$_2$, there is a clear difference in the propensity to induce a SOI between monolayer and bulk. Bulk TMDs induce SOI in graphene that is more than one order of magnitude smaller than monolayer ones. Thus we conclude that monolayer WSe$_2$ and WS$_2$ can induce the strongest SOI in graphene. 

As a universal behavior, we found that in all systems the induced SOI is predominantly $z \rightarrow -z$ symmetric, composed of KM or VZ SOI. The analysis of the spin relaxation mechanism indicates that the KM SOI plays an important role, especially close to the Dirac point.

While more experimental and theoretical work is still necessary for a deeper understanding, our experimental findings offer new insights on SOI in graphene produced by TMDs, and provide important information for application to spintronics and topological physics.   

\section{\label{sec:level1}Acknowledgements\protect\\ }
We gratefully acknowledge very useful discussions with V. Fal'ko, A. Meszaros, A. W. Cummings, S. Roche, P. Makk, S. Zihlmann and C. Sch\"{o}nenberger. This project is financially supported in part by the Marie
Sklodowska Curie Individual Fellowships (H2020-MSCAIF-2014-659420); the ANR Grants DIRACFORMAG (ANR-14-CE32-003), MAGMA (ANR-16-CE29-0027-02), and JETS (ANR-16-CE30-0029-01), the Overseas Research Fellowships by the Japan Society for the Promotion of Science, the CNRS and the award of a Royal Society University Research Fellowship by the UK Royal Society, the EPSRC grant EP/M022250/1 and the EPSRC-Royal Society Fellowship Engagement Grant EP/L003481/1. M.Q. Zhao and A. T. C. Johnson are supported by the US NSF grant EFRI 2-DARE 1542879.

\end{document}